\documentclass[12pt]{iopart}

\usepackage[dvips]{graphicx}

\usepackage[numbers,compress]{natbib}
\def\newblock{\hskip .11em plus .33em minus .07em}
\usepackage[colorlinks]{hyperref}
\usepackage{color}
\usepackage{setstack}
\usepackage{iopams}

\def\wigwid{1}
\def\distwid{0.6}

\def\ket#1{|#1\rangle}
\def\expct#1{\langle#1\rangle}
\def\gket{\ket{\mathrm{g}}}
\def\eket{\ket{\mathrm{e}}}
\def\omegaa{\omega_{\mathrm{\scriptscriptstyle{A}}}}
\def\omegac{\omega_{\mathrm{\scriptscriptstyle{C}}}}

\begin{document}

\title{Reflection of a Particle from a Quantum Measurement}
\author{Jonathan B. Mackrory$^1$, Kurt Jacobs$^2$, Daniel A. Steck$^1$}
\address{$^1$ Department of Physics and Oregon Center for Optics, 1274 University of Oregon, Eugene, OR 97403-1274}
\address{$^2$ Department of Physics, University of Massachusetts at Boston, 100 Morrissey Blvd, Boston, MA 02125, USA}


\begin{abstract}
We present a generalization of continuous position measurements that accounts for a spatially inhomogeneous measurement strength. This describes many real measurement scenarios, in which the rate at which information is extracted about position has itself a spatial profile, and includes measurements that detect if a particle has crossed from one region into another.  We show that  such measurements can be described, in their averaged behavior, as  stochastically fluctuating potentials of vanishing time average.  Reasonable constraints restrict the form of the measurement to have degenerate outcomes, which tend to drive the system to spatial superposition states. We present the results of quantum-trajectory simulations for measurements with a step-function profile (a ``which-way'' measurement) and a Gaussian profile.  We find that the particle can coherently reflect from the measurement region in both cases, despite the stochastic nature of the measurement back-action. In addition, we explore the connection to the quantum Zeno effect, where we find that the reflection probability tends to unity as the measurement strength increases.  Finally, we discuss two physical realizations of a spatially varying position measurement using atoms. 
\end{abstract}
\maketitle
\makeatletter
\renewcommand\tableofcontents{\section*{\contentsname}\@starttoc{toc}}
\makeatother
\tableofcontents
\section{Introduction}
Position measurements strike at the heart of what distinguishes quantum mechanics from classical mechanics.  Within classical mechanics, it is possible to localize particles with arbitrarily fine precision for all times. In quantum mechanics, the back-action from the measurement disturbs the particle and implies that there are fundamental limits to measurement precision---limits that are being approached in current experiments~\cite{Rocheleau10,Murch08}.  

The usual formalism of projective measurements fails when applied to position measurements: if a particle is reduced to an eigenstate of position it has an infinite momentum uncertainty and hence infinite energy.  Position measurements are best described by considering the ``weak'' limit where the observer continuously extracts information about the particle's location at a given rate; in a small time interval only a small amount of information is extracted.   

Continuous measurements for quantum systems are usually phrased in the language of stochastic master equations, where the system's density matrix is updated based on random measurement outcomes~\cite{Belavkin87}.  The stochastic master equation can also be derived from a physical model for the interaction of the system with the measurement apparatus~\cite{Wiseman93}. The first model specifically for a continuous position measurement accounted for the ensemble-averaged evolution of the density matrix~\cite{Caves87}.  This has been developed further into the modern description in which stochastic master equations (quantum trajectories) are used to continuously condition the quantum state based on the output of a classical detector that is correlated with the system's position~\cite{Gagen93}.  The observer can equivalently monitor the status of a large bath with which the system interacts. The scenario in which the resonance fluorescence of an atom is monitored to gather position information is one example of this~\cite{Holland96, Jacobs06}.   A similar approach is used to treat measurements of the position of an atom  interacting with the field mode of an optical cavity~\cite{Doherty98, Hood00, Doherty99} .  

Continuous measurements have an important role in quantum control and in the transition to the classical limit of quantum mechanics. Continuous \textit{position} measurements in particular can be employed in feedback control loops for cooling quantum systems~\cite{Doherty99}. This has been applied to atoms in cavities~\cite{Fischer02, Steck04, Steck06, Kubanek09}, trapped ions~\cite{Steixner05, Bushev06}, and nanomechanical resonators~\cite{Hopkins03}.  In addition to seeking to  control quantum systems, continuous position measurements also provide one path through the quantum-to-classical transition. In particular, continuously monitored quantum systems are able to exhibit chaotic behavior~\cite{Bhattacharya05, Habib06}, in contrast to closed quantum systems. 

Considering the importance of position measurements, it is necessary to develop the theory to account for realistic constraints applicable to any experimental realization of a position measurement.  In particular, in any real position measurement the particle can only be detected within a limited region.  If the particle leaves this region, the observer gains no further knowledge of the particle's position.  This can be modeled by a space-dependent coupling of the particle to a bath, such as the radiation field. When the strength of the coupling to the bath, and thus the \textit{measurement strength}, is itself a function of the position, it is, in fact, \textit{this function} of the position that becomes the measured observable. The coupling correlates the bath with that function of the particle's position, and so monitoring the bath allows the observer to gather information about that function.

The average dynamics of the measured particle (that is, the motion of the particle averaged over all the possible measurement results) can be reproduced by a fluctuating potential with the same position dependence as the measurement. Thus, as far as the average motion is concerned, the measurement acts like a stochastic force. This fluctuating potential provides an alternate, intuitive physical picture for understanding the measurement back-action.  We expand on this unitary ``unraveling'' of the average  motion in Sec.~\ref{sec:stochastic_potential}.  

It has been shown previously by a number of authors that a measurement that determines whether a particle is on one side of a dividing line or the other can exclude the wave-function from the region in which the particle is initially absent~\cite{Allcock69_1, Allcock69_2, Allcock69_3, Misra77, Facchi01, Facchi04, Exner05, Koshino05, Echanobe08}. This causes the particle to reflect from the dividing line, and is due to the quantum Zeno effect. In our case this corresponds to a measurement with the following two properties: 1) the measurement strength takes one of just two values, and makes a jump from one value to the other at the dividing line (the measurement strength is a step function); 2) the measurement strength (equivalently, the size of the step) is much greater than the kinetic energy of the incoming particle, suitably scaled. In this case the bath exerts a large back-action on the particle, providing the momentum transfer required for the reflection. While this Zeno effect has been discussed previously, we examine it here to connect it with the stochastic master equation formalism. We do this by deriving an analytic expression for reflection probability in Sec.~\ref{sec:reflection}. We also perform numerical simulations of a measurement with a step-function profile, for comparison to a localized measurement with a Gaussian profile.

A position measurement can be physically realized by monitoring the resonance fluorescence of an atom interacting with a resonant light field.  The profile of the light becomes the measurement function, and the measurement strength is proportional to the overall intensity.  As a consequence of the aforementioned effects, the atom will \textit{coherently reflect} from the resonant light field for a sufficiently large intensity, despite the absence of a mean dipole force, and despite the stochastic nature of resonant atom--light interactions that tend to heat the atom.  
This phenomenon, besides being somewhat counterintuitive, has implications in situations where atoms encounter localized, resonant, optical-pumping fields---as occurs, for example, in implementations of one-way barriers for atoms~\cite{raizen2005, ruschhaupt2004, price2008, thorn2008}.  We present further details on this realization in Sec.~\ref{sec:physical realization}, along with another example of an atom interacting with an off-resonant cavity.

We begin in the next section by deriving the stochastic master equation that describes a continuous position measurement with a spatially varying measurement strength, and elucidate some of its key properties. In Sec.~\ref{sec:simulations} we perform simulations of a particle incident on measurements with two kinds of spatial profiles, showing the behavior of the particle on individual realizations of the measurement, which may involve either reflection or transmission, as well as the ensemble-averaged behavior. In Sec.~\ref{sec:reflection} we discuss the quantum Zeno effect, in Sec.~\ref{sec:physical realization} we present two physical realizations of a spatially varying measurement, and in Secs.~\ref{sec:analogies} and~\ref{sec:outlook} we finish with some concluding remarks. 

\section{Equations of Motion}
\label{sec:eqn of motion}
\subsection{Derivation}
In this section we will derive the equation of motion describing a spatially varying position measurement.   Our measurements are usually made by monitoring a large bath that interacts irreversibly with the system.  The position measurement arises from a spatially-dependent potential that couples the particle to the bath.  The bath then becomes correlated with a real function of the particle's position $\mu(x)$ and we can distinguish different positions by how strongly they interact with the bath.   Although $\mu$ is real, we will keep our notation general because later we will add a complex local-oscillator amplitude to $\mu$.  We will start our derivation from the positive-operator-valued measure for a measurement with two outcomes in each infinitesimal time-step, $dt$ (note that our treatment and superoperator notation parallels that of Ref.~\cite{Wiseman10QMAC}).   We will refer to the two outcomes respectively as ``no detection,'' and ``detection''. (We choose this nomenclature due to the formal similarity to photodetection, though these outcomes are not necessarily tied to a photodetector.)  The measurement operators that describe the two outcomes are 
\begin{eqnarray}
\Omega_0 & =  1 - \frac{i}{\hbar}H\, dt -\kappa\mu^\dag(x)\mu(x)\,dt\\
\Omega_1 & = \sqrt{2\kappa\, dt}\,\mu(x),
\end{eqnarray}
where $H=p^2/(2m)+V(x)$ is the free Hamiltonian for the system, $\kappa$ is the measurement strength, $\Omega_0$ corresponds to ``no detection'',  and $\Omega_1$ to a ``detection''.   The detection outcome occurs with probability $\mathrm{Tr}[\Omega^\dag_1\Omega_1\rho] = 2\kappa\langle\mu^\dag(x)\mu(x)\rangle\,dt.$    The evolution that occurs between detections is due to $\Omega_0,$ 
\begin{eqnarray}
\rho \rightarrow & \frac{\Omega_0\rho\Omega_0^\dag}{\mathrm{Tr}[\Omega_0^\dag\Omega_0\rho]} = \rho -\frac{i}{\hbar}[H,\rho]\,dt - \kappa\mathcal{H}[\mu^\dag(x)\mu(x)]\rho \,dt,
\end{eqnarray}
where the $\mathcal{H}$ super-operator is defined as 
\begin{equation}
\mathcal{H}[c]\rho  = c\rho + \rho c^\dag - \langle c+c^\dag \rangle \rho.  
\end{equation}
If detection occurs the density matrix changes according to
\begin{eqnarray}
\rho \rightarrow  \frac{\Omega_1\rho\Omega_1^\dag}{\mathrm{Tr}[\Omega_1^\dag\Omega_1\rho]} =  \frac{\mu(x)\rho\mu^\dag(x)}{\mathrm{Tr}[\mu^\dag(x)\mu(x)\rho]}.
\end{eqnarray}
We can represent all of this evolution in a single stochastic master equation (SME):
\begin{equation}
d\rho = -\frac{i}{\hbar}[H,\rho]\,dt -\kappa\mathcal{H}[\mu^\dag(x)\mu(x)]\rho \,dt+ \mathcal{G}[\mu(x)]\rho \,dN,
\label{eq:position jump ME}
\end{equation}
where
\begin{eqnarray}
\mathcal{G}[c]\rho = \frac{c\rho c^\dag}{\mathrm{Tr}[c^\dag c\rho]}-\rho.
\end{eqnarray}
Here, $dN$ is a Poisson process~\cite{JacobsSP, GardinerSM}, where $dN=1$ with probability $2\kappa\langle \mu^\dag(x)\mu(x)\rangle \,dt$ and is zero otherwise.
If we mix in a ``local oscillator'' with this signal, then we pass over to a master equation similar in form to the usual position-measurement master equation~\cite{Jacobs06}.  This corresponds to the transformation
\numparts
\begin{eqnarray}
\mu & \rightarrow \mu+\frac{\alpha}{\sqrt{2\kappa}}\label{eq:homodyne transform1}\\
H &\rightarrow H -\frac{i\hbar\sqrt{2\kappa}}{2}(\alpha^*\mu-\alpha \mu) \label{eq:homodyne transform2},
\end{eqnarray}
\endnumparts
where $\alpha=|\alpha|e^{i\phi}$ is the complex amplitude of the local oscillator, which leaves the unconditioned evolution  (that is, the evolution averaged over all the possible measurement results) unchanged.  For large $|\alpha|$, we can pass over to the white noise limit.  When the detection rate becomes fast on the time scale of the particle's motion, the central limit theorem allows us to replace $dN$ by a mean drift and Gaussian fluctuations,
 \begin{eqnarray}\label{eq:white noise}
dN & = \left<\hspace{-0.3em} \left<\frac{dN}{dt}\right> \hspace{-0.3em} \right>dt + \sqrt{\left< \hspace{-0.3em}\left<\frac{dN}{dt}\right> \hspace{-0.3em}\right>}\,dW,
\end{eqnarray}
where double angle brackets denote the ensemble average for the random variable, and $dW$ is Ito white noise, with $\left<\!\left<dW\right>\!\right>=0$ and $dW^2=dt$~\cite{JacobsSP, GardinerSM}.   We apply the transformations in Eqs.~(\ref{eq:homodyne transform1}), (\ref{eq:homodyne transform2}),  and the white noise limit of Eq.~(\ref{eq:white noise}), to the jump master equation (\ref{eq:position jump ME}).  In the large $|\alpha|$ limit, the resulting white noise master equation is 
\begin{equation}\label{eq:SME_measurement}
d\rho = -\frac{i}{\hbar}[H,\rho] \,dt +2\kappa\mathcal{D}[\mu(x)]\rho \,dt + \sqrt{2\kappa}\mathcal{H}[e^{-i\phi}\mu(x)]\rho \,dW,
\end{equation}
 where the $\mathcal{D}$ superoperator is defined as
\begin{equation}
\mathcal{D}[c]\rho = c\rho c^\dag - \frac{1}{2}\left(c^\dag c\rho + \rho c^\dag c\right).
\end{equation}
For $\phi=0$, Eq.~(\ref{eq:SME_measurement}) is simply the standard stochastic master equation that describes the measurement of the observable $O = \mu(x)$~\cite{Jacobs06}.  Thus the case $\mu(x)=x$ recovers the usual master equation for a measurement of position.  The continuous stream of measurement results, often referred to as the \textit{measurement record}, is given by 
\begin{equation}
    dr =   \langle \mu(x) \rangle dt + \frac{dW}{\sqrt{8 \kappa}} , 
\end{equation}
where $dr$ is the result in the time interval $dt$.  

The unconditioned dynamics are given by the master equation 
\begin{equation}\label{eq:ME_uncond}
\partial_t\rho = -\frac{i}{\hbar}[H,\rho] - \kappa [\mu(x),[\mu(x),\rho]],
\end{equation}
where we have taken an ensemble average over all possible noise realizations.

\subsection{Stochastic Potential}
\label{sec:stochastic_potential}
An alternative interpretation is suggested if we choose a different phase for the local oscillator.  If we instead choose $\phi =\pi/2$, the master equation becomes 
\begin{eqnarray}\label{eq:SME_potential}
d\rho &= -\frac{i}{\hbar}[H\,dt+\sqrt{2\kappa}\hbar\mu(x)dW,\rho]  +2\kappa\mathcal{D}[\mu(x)]\rho \,dt,\\
 & = -\frac{i}{\hbar}[H\,dt+\sqrt{2\kappa}\hbar\mu(x)\circ dW,\rho]\label{eq:potential_strat},
\end{eqnarray}
where the first equation is in Ito form, and in the second equation ${}\circ dW$ denotes Stratonovich white noise~\cite{JacobsSP, GardinerSM}.  This phase choice corresponds to measuring the other bath quadrature \cite{Milburn96}, yielding information about the measurement backaction rather than position.  The effect of the bath here is formally equivalent to a stochastic potential.  The spatial dependence $\mu(x)$ of the coupling to the bath determines the profile of the stochastic potential.  The term in $\mathcal{D}[\mu(x)]$ is the Ito correction to the Stratonovich fluctuating potential. When averaged over all noise realizations, this master equation recovers the unconditioned evolution given by Eq.~(\ref{eq:ME_uncond}).

This view is particularly apparent if we look at the input-output formulation of a homodyne measurement \cite{Goetsch94}.  The measurement arises from a linear coupling between the observable and the bath, 
\begin{equation}
H_\mathrm{SB}\,dt =\mu(x)(dB_{\mathrm{in}}+dB^\dag_{\mathrm{in}}),
\end{equation}
where $dB_{in}$ is a white noise bath operator due to the sum over bath modes, all evolving at different frequencies~\cite{Gardiner85}.  This interaction term correlates the state of the bath with $\mu(x)$.  The key observation is that since the interaction Hamiltonian is proportional to $\mu(x)$, the noise in the bath drives the system through $\mu(x)$.  The bath therefore exerts a stochastic force on the particle whenever $\mu(x)$ is inhomogeneous.  Note that \textit{any} measurement of $\mu(x)$ that is mediated by a bath \textit{must} have an interaction of this form~\cite{stochastic_footnote}.  This in turn implies that there \textit{must} be a stochastic force of the same form.  As a concrete example, a two-level atom interacting with a resonant laser field feels the stochastic dipole force due to spontaneous emission~\cite{Dalibard85}.

\subsection{Effects of Measurement}

 We can gain insights into the dynamics induced by a spatially varying position measurement  by examining the lowest few terms in the Taylor expansion of $\mu(x)$ about the current mean position $\langle x\rangle$.  If $\mu(x)$ is constant then the post-measurement state is completely unaffected, and the measurement does nothing.  

The linear term in the Taylor expansion acts as a standard position measurement.  The measurement can then be approximated by 
\begin{equation}\label{eq:linear measurement}
\sqrt{2\kappa} \mu(x) \approx \sqrt{2\kappa}\mu'(\langle x\rangle)(x-\langle x\rangle),
\end{equation}
which defines the effective local measurement strength, where $\mu'(x) \equiv d\mu/dx$.  The constant $\langle x\rangle$ term again has no effect. This linear approximation is valid if the particle is well-localized on the scale of the measurement function.  This relation provides the link between a general, inhomogeneous measurement of position and the usual, linear position measurement.   

The quadratic term in the Taylor expansion is important at the maxima and minima of the measurement function where the linear term vanishes. The effective measurement is 
\begin{equation}
\sqrt{2\kappa}\mu(x) \approx \sqrt{2\kappa}\mu''(\langle x\rangle)(x-\langle x\rangle)^2,
\end{equation}
which drives the state to spatial superpositions since the measurement result cannot distinguish $\langle x\rangle + x$ from $\langle x \rangle - x$~\cite{Jacobs09}.    Once the components of the superposition move away from the extremum, the local measurement for each component becomes linear.  Each component will then evolve under their respective effective measurements.

\begin{figure}
\includegraphics[width=\columnwidth]{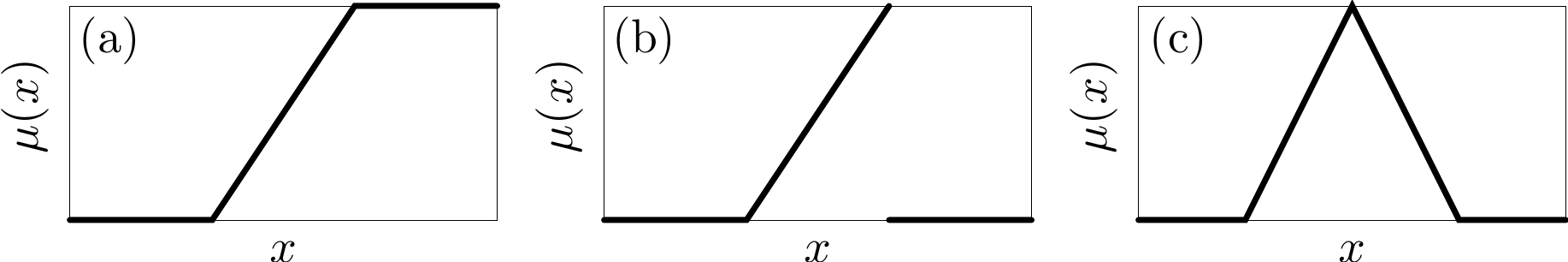}
\caption{Possible measurement functions realizing a position measurement within a bounded region. The local measurement strength is proportional to the derivative of $\mu(x)$.  (a) Measurement function with different asymptotic values. (b) Measurement function with a discontinuous return to zero. (c) Continuous measurement function with same asymptotic values.}
\label{fig:measurement_func}
\end{figure}

\subsection{Form of the Measurement Function}

We now derive the conditions that the measurement function $\mu(x)$ must satisfy for it to correspond to a position measurement that acts only in a limited region of space.  First we expand the measurement terms in the master equation (\ref{eq:SME_measurement}) in the position basis, with $\rho(x,x')=\langle x|\rho|x'\rangle$:
\begin{eqnarray}\label{eq:ME position basis }
  d\rho(x,x') =& -\kappa[\mu(x)-\mu(x')]^2\rho(x,x')\,dt\nonumber \\
&+\sqrt{2\kappa}\left[\mu(x)+\mu(x')-2\langle \mu\rangle\right]\rho(x,x')\,dW.
\end{eqnarray}
The leading term leads to decay of coherence between positions $x$ and $x'$ where $\mu$ differ.  If we enforce $\mu(x)=\mu(x')$, then the measurement cannot distinguish between $x$ and $x'$.  This implies that $\mu(x)$ should be constant outside the measurement region.  

Inside the measurement region, the measurement function $\mu(x)$ should be linear to act as a standard position measurement. However, if $\mu(x)$ does not return to the same value on either side of its linear section (as in Fig.~\ref{fig:measurement_func}(a)), then the measurement would collapse superpositions that are on either side of, and completely outside the linear measurement region. 
This is not what one normally means when restricting a measurement to a particular region.  For example, a photodetector of some finite size placed well within the arms of an optical interferometer should not collapse the fringes by giving which-way information, but this is precisely the case for the function in Fig.~\ref{fig:measurement_func}(a). Thus, for a measurement to truly act only within some bounded region, the function must  take on the same value everywhere outside that region.
We may take this value to be zero, since shifting $\mu(x)$ by a constant has no effect on the dynamics.  

Generally speaking, $\mu(x)$ will be the profile of some continuous (e.g., laser) field.  Thus, a discontinuous return of the measurement function to the external value, as in Fig.~\ref{fig:measurement_func}(b), is at best an idealization of a more physical, continuous measurement function. Along the same lines, the measurement should also be finite in extent.  Measurement functions of the type in Fig.~\ref{fig:measurement_func}(a) must therefore also be idealizations of measurement functions that are eventually zero, though possibly only returning to zero far away from the location of the measured particle.

The strictly physical form of the measurement function must therefore be continuous and return to the same constant value on both sides of the measurement region, as shown in Fig.~\ref{fig:measurement_func}(c).  In particular, this means that there will \textit{always} be indistinguishable positions $x_1,\  x_2$, where $\mu(x_1) = \mu(x_2)$ within the measurement region, corresponding to the same measurement outcome.  At any given instant, the measurement cannot distinguish between the positions $x_1,\ x_2$, even if the measurement is not symmetric on either side of the extremum, for example if $\mu'(x_1)\ne \mu'(x_2)$.  The measurement will thus create a superposition.  However, if we also account for the Hamiltonian evolution then the backaction from an asymmetric measurement can give the components of the superposition differing momenta, and hence different positions at later times.  Hence over an extended period of time the measurement could distinguish between the two positions. 
Likewise, we have argued that for a single measurement channel, a strictly physical measurement function must result in positions that are indistinguishable.  However, multiple position-measurement channels---each with their own ambiguities---may be combined to remove \textit{all} the ambiguities.  These are examples of combining multiple measurements to gain information that cannot be gleaned from either measurement separately.  A more familiar example is that information from position measurements at two different times may be combined to obtain information about momentum.

\subsection{Mean Momentum and Kinetic Energy}
We can derive equations of motion for the two lowest-order moments of the momentum probability density based on Eq.~(\ref{eq:SME_measurement}).  If we assume the particle Hamiltonian is given by 
\begin{equation}
H = \frac{p^2}{2m} + V(x),
\end{equation}
we obtain
\begin{eqnarray}
d\langle p\rangle  = &-\langle \partial_xV\rangle \,dt +\sqrt{2\kappa}[\langle \mu p+p\mu\rangle -2\langle \mu \rangle\langle p\rangle]\,dW \label{eq:mean p}\\
d\langle p^2\rangle  =& -\langle p\partial_xV+\partial_xV p\rangle \,dt+2\hbar^2\kappa\langle (\partial_x\mu)^2\rangle \,dt \nonumber \\
&+\sqrt{2\kappa}[\langle \mu p^2+p^2\mu\rangle -2\langle \mu \rangle\langle p^2\rangle]\,dW.
\end{eqnarray}
In an ensemble average $d\!\left<\!\left< p\right>\!\right>=-\langle \partial_xV\rangle\, dt$, so the \textit{average} momentum is not changed by the measurement.  For a free-particle ensemble, the average momentum is conserved---during a single trajectory, the measurement can exert a force, but the force vanishes on average.  The measurement back-action enters as diffusion in the ensemble-averaged moment evolution,
\begin{equation}
  d\!\left<\!\left< p^2\right>\!\right>= 2\hbar^2\kappa\langle (\partial_x\mu)^2\rangle \,dt,
  \label{kappadiffusioncoeff}
\end{equation}  
with momentum diffusion coefficient $D_\mathrm{p} = 2\hbar^2\kappa\langle (\partial_x\mu)^2\rangle$.
This reflects that it is the \textit{gradient} in $\mu(x)$ that gives relative position information.   

\section{Simulations}
\label{sec:simulations}

We now present some quantum-trajectory simulations that reveal some of the novel dynamics that occur under inhomogeneous position measurements.  In the simulations, the particle starts in a minimum-uncertainty, Gaussian state with mean position  $\langle x\rangle=x_0$,  mean momentum $\langle p\rangle=p_0$ and root-mean-square width $\sigma_x$.  It is incident on a measurement function $\mu(x)$ from outside the measurement region, from the left-hand side.  Since the atom starts in a pure state we initially have maximal knowledge about the atom.  Our results focus on the case when $p_0 /\sigma_p\gg 1$, which corresponds to a well-defined momentum.  In the limit of small $\kappa$, the particle simply passes through the measurement region with mild heating, so we also restrict ourselves to the regime of strong measurements, $\kappa > p_0^2/(2m\hbar)$.

Our simulations use a split-operator Fourier method to propagate the stochastic Schr\"odinger equation associated with Eq.~(\ref{eq:SME_measurement}).  
We have set $\hbar=m=1$ throughout our presentation here. 
We visualize the evolution of the particle under the measurement using the Wigner function for the particle~\cite{Schleich_phase_space}.  

\subsection{Step function}\label{section:step-function}

\begin{figure*}
\centering
\includegraphics[width=\wigwid\textwidth]{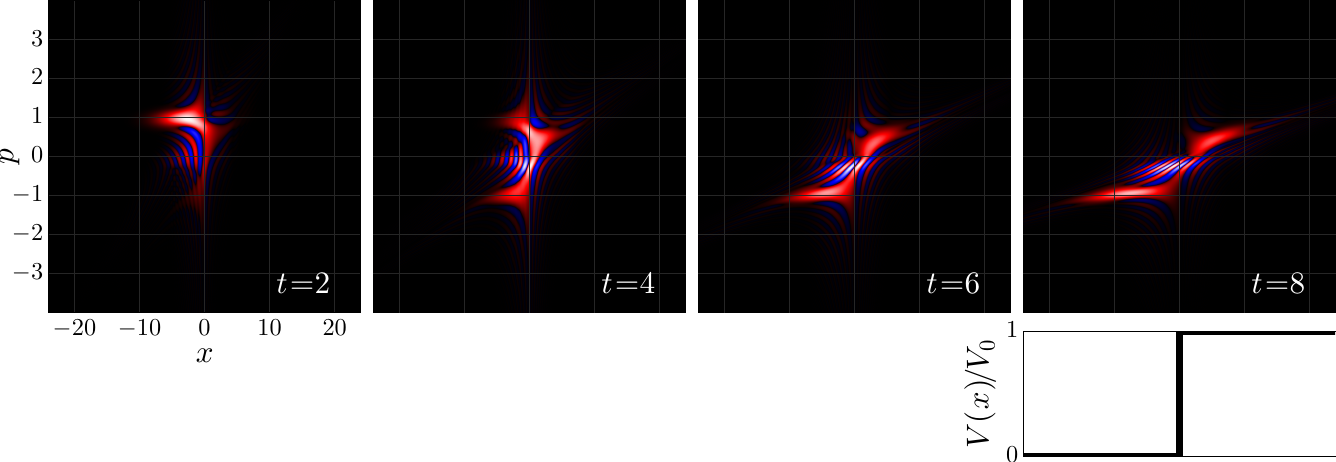}
\caption{Wigner functions for a particle incident on a static potential step with height $V_0 =0.5$, and without measurement.  The wave packet has initial momentum $p_0=1$, width $\sigma_x=5,$ and initial position $x_0=-3\sigma_x$. Time is measured in units of $\sigma_x/p_0$. Red values are positive, blue values are negative, and black is zero. The corresponding animation is included in the supplementary data. \label{fig:step_potential}}

\includegraphics[width=\wigwid\textwidth]{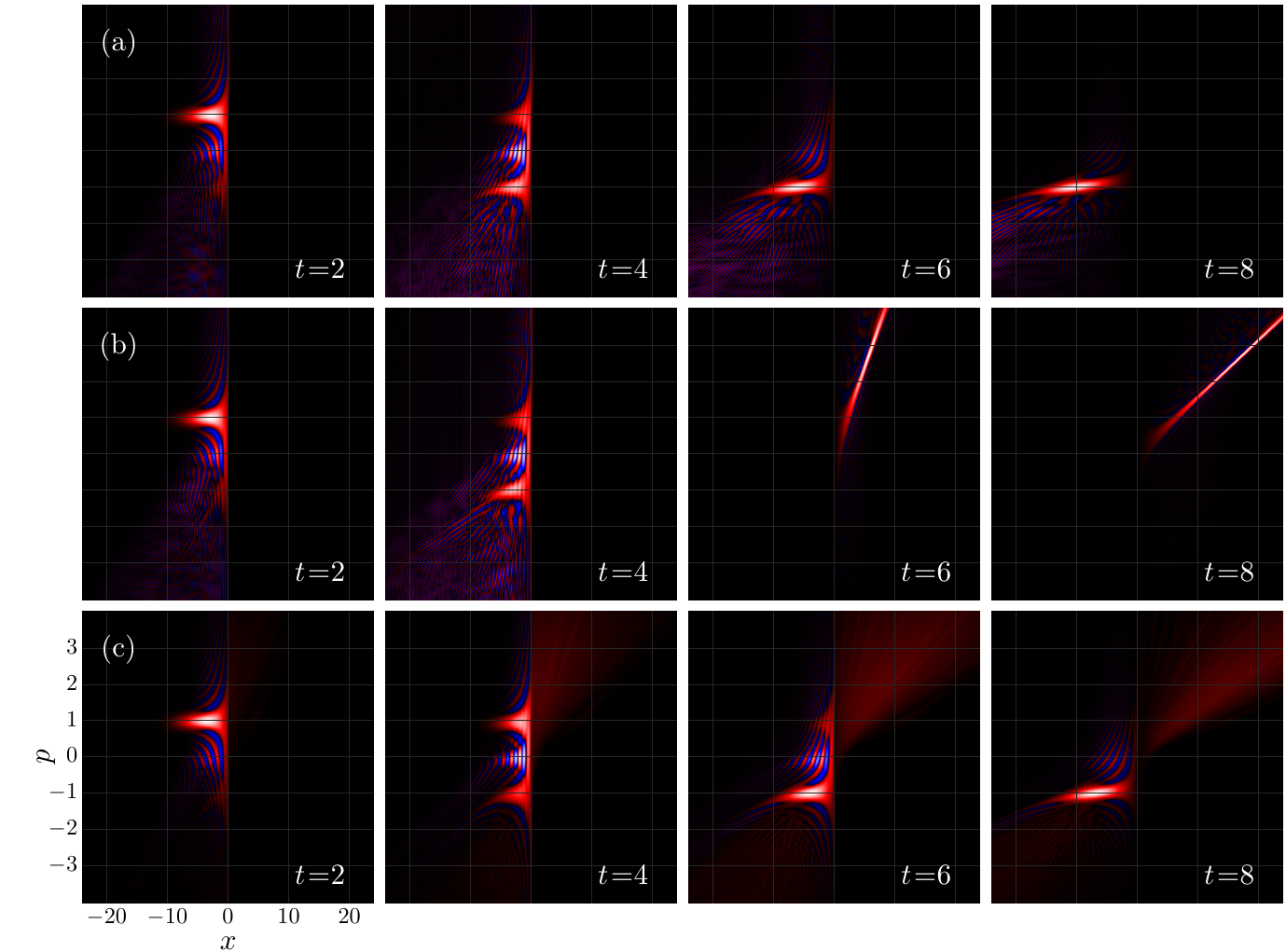}
\caption{Wigner functions for a particle incident on a step-function measurement profile 
$\mu(x)=\theta(x)$ with $\kappa=5$; the profile is the same as the potential in Fig.~\ref{fig:step_potential}.  The trajectories are generated with a white-noise-measurement unraveling, as given by the Eq.~(\ref{eq:SME_measurement}). 
The initial conditions are the same as in Fig.~\ref{fig:step_potential}.   
(a) Single trajectory with reflecting measurement outcome.
(b) Single trajectory with transmitting measurement outcome.
(c) Ensemble average over measurement realizations.
Red values are positive, blue values are negative, and black is zero. The corresponding animations are included in the supplementary data.
\label{fig:step_measurement}}
\end{figure*}

\label{sec:step}
We will first consider a step measurement function
\begin{equation}
\mu(x) = \theta(x) = \bigg\{\begin{array}{ll} 
1 & \quad x > 0\\
0 & \quad x < 0,
\end{array}
\end{equation}
which is simple and analytically tractable.  It also affords a comparison with a familiar example from textbook quantum mechanics: the potential step.  This corresponds to the limit of an arbitrarily large gradient for a continuous measurement.  This limit is unphysical, strictly speaking, but it is useful as an idealization of a strong measurement over a small region. 

Fig.~\ref{fig:step_potential} shows the temporal evolution of a  wave packet incident on a static potential step $V(x) = V_0 \theta(x)$, with no measurement, as a context for understanding the measurement-driven evolution.  The evolution can be described by the Schr\"odinger equation, or by setting $\kappa=0$ in the measurement master equation (\ref{eq:SME_measurement}), and including the potential $V(x)$.  The height of the potential barrier in this case is smaller than the kinetic energy of the wave packet, so the particle can transmit over the barrier.  The wave packet splits into two pieces, reflected and transmitted.  The reflected wave packet has exactly the opposite momentum of the inbound wave packet.  However, the transmitted wave packet is slower, since the particle lost energy climbing the potential step.  The striated red and blue regions denote interference.   The fringes between the reflected and transmitted wave packets show that they represent a coherent superposition.  Note that the fringes between the two coherent components are oriented along the direction between them.  This corresponds to the fact the fringes would yield an interference pattern  in the appropriate marginal probability distributions \cite{Schleich_phase_space}.  Missing fringes indicate a \textit{classical} mixture of the two states.  

Fig.~\ref{fig:step_measurement} shows the evolution of a wave packet incident on a measurement step $\mu(x)=\theta(x)$, but for a free particle with no explicit potential.  The step measurement distinguishes whether the particle is on the left- or the right-hand side of the origin, but it does not resolve different positions on a given side.  This means that superpositions confined to one side are unaffected by the measurement,  while superpositions on different sides of the origin will collapse.  This measurement is analogous to a ``which-way'' measurement in that it has only two outcomes: left or right.  The two possible measurement  outcomes correspond to reflection or transmission of the wave packet.  

Fig.~\ref{fig:step_measurement}(a) shows a single trajectory where the wave packet reflects from the measurement.  Initially, we know the particle is to the left of and propagating towards the step in the measurement function.  As the wave packet evolves, we never measure it to be on the other side, and this measurement outcome effectively forces the wave packet to reflect.  The momentum transfer for the reflection comes from the back-action of the measurement.  As noted in Sec.~\ref{sec:stochastic_potential}, a spatially dependent measurement implies a spatially varying back-action on the system.  

\begin{figure*}
\centering
\includegraphics[width=\distwid\columnwidth]{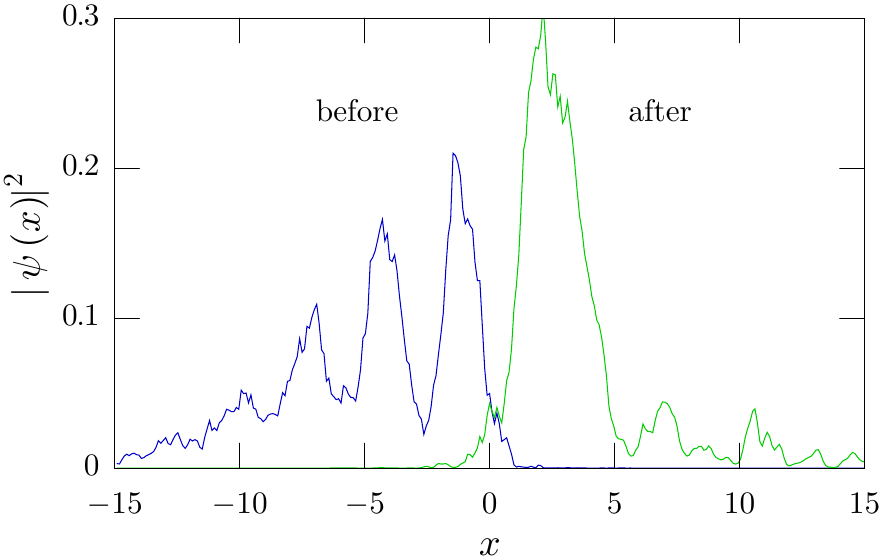}
\includegraphics[width=\distwid\columnwidth]{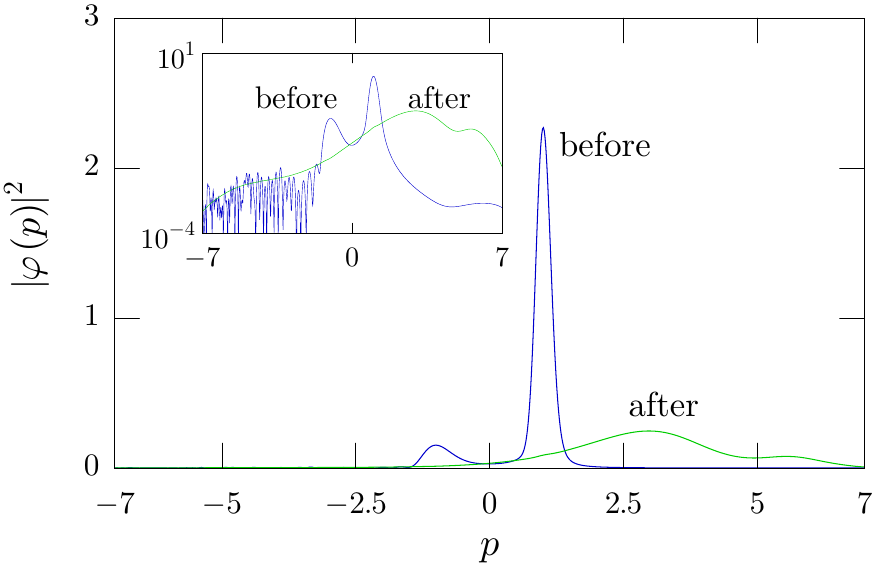}
\caption{Position and momentum distributions for a particle  before and after it transmits through the step measurement.  The initial conditions are the same as in Fig.~\ref{fig:step_potential}.  The time elapsed between ``before'' and ``after'' is $\Delta t = 0.4\kappa^{-1}$. Note that this is not the same trajectory as in Fig.~\ref{fig:step_measurement}(b), but it displays a similar behavior.  Inset: log plot of the same momentum distribution. }
\label{fig:distributions}
\end{figure*}

Fig.~\ref{fig:step_measurement}(b) shows an example where the particle transmits through the measurement step.  The wave packet starts to reflect, but still has a small evanescent tail that penetrates the $x>0$ region.  The interference between the reflected and initial wave packets, before transmission occurs, can be seen in Fig.~\ref{fig:distributions}.  Random measurement outcomes consistent with the particle being in the $x>0$ region are rare at first since the evanescent tail represents a small probability.  Once they occur, however, these outcomes can increase the amplitude of that part of the wave function.  This can lead to a runaway process where future measurement results are more likely to indicate the particle is in the $x>0$ region, and thereby collapse the particle entirely into this region.  This process occurs continuously on a time scale $\kappa^{-1}$.  Since the measurement time scale is faster than that of the motion in the strong-measurement limit, the particle is localized in the small, evanescent tail before it can propagate away.  The particle will now have a broad, positive momentum distribution and will propagate away from the step, as shown after transmission occurs in Fig.~\ref{fig:distributions}.  The transmission process is incoherent: the state of the particle is entangled with the state of the bath, which destroys any interference with the incident wave packet.  Since transmission occurs at random times, the transmitted wave packets will have different phases and propagate away at different times.  

Fig.~\ref{fig:step_measurement}(c) shows an ensemble average of $10^4$ trajectories for particles incident on the step measurement.  Note that the specular component of the reflection survives the ensemble average, while other features related to the measurement noise are washed out in the average.  The transmitted parts show no coherence with one another or with the reflected parts.  The incoherent parts of both reflected and transmitted wave packets show heating from the measurement back-action.   

Surprisingly, part of the reflection is coherent.  It is possible to see interference between the reflected and incident wave packets, even though the reflection is caused by the back-action of a measurement,
which conventionally leads to diffusion or heating.  The interference is marked by the alternating positive and negative regions centered at $p=0$ in the Wigner function.  The interference fringes have the same phase for each member of the ensemble---that is, they are the same for each possible, random outcome of the measurement---producing the coherent reflection.

The measurement-induced reflection is analogous to interaction-free measurements~\cite{Kwiat99}.  In those experiments, light detects the presence of an absorber in the arms of a polarization interferometer without ever interacting with the absorber.  An optical example of an interaction-free measurement closely related to the reflection phenomenon here is that of a high-finesse Fabry--Perot cavity~\cite{Paul96, Tsegaye98}. An empty cavity transmits light almost perfectly, but when an absorber is placed in the cavity to ``detect'' the light, a large fraction of the light instead reflects from the cavity, although the reflected part has not interacted with the absorber.

\subsection{Gaussian measurement}
\label{sec:Gaussian}

We now consider the case of a Gaussian measurement function.  This is an even function of position, so we expect this measurement to tend to drive the particle towards spatial superposition states.  In addition, if we consider the resonance-fluorescence scenario, a laser probe with a Gaussian profile would realize this Gaussian measurement.  The measurement function is
\begin{equation}
\mu(x) = \exp\left[-\frac{x^2}{2\sigma_\mu^2}\right],
\end{equation}
where $\sigma_\mu$ sets the width of the measurement.  Note that we take $\mu(x)$ to be dimensionless, which is important to keep in mind when comparing to standard measurements $\mu(x) = x$, which makes sense only when $x$ is taken to be dimensionless.  

\begin{figure*}
\centering
\includegraphics[width=\wigwid\textwidth]{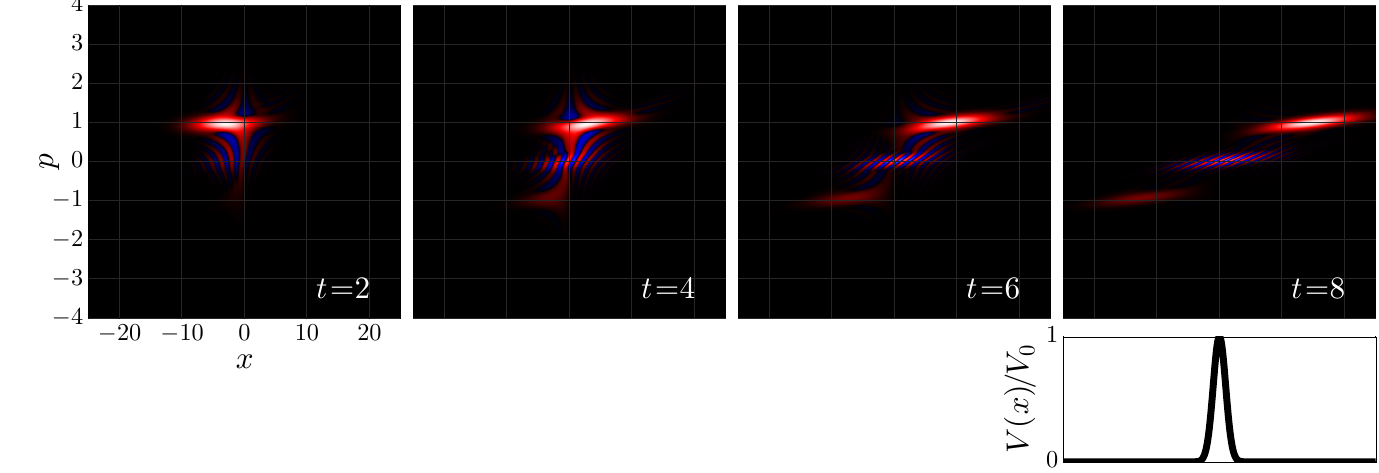}
\caption{Wigner functions for particle incident on a static Gaussian potential, with potential width $\sigma_\mu=1$ and height $V_0=0.25$, without measurement.  The wave packet has initial momentum $p_0=1$, width $\sigma_x=5,$ initial position $x_0=-3(\sigma_x+\sigma_\mu)$. Time is measured in units of $(\sigma_x+\sigma_\mu)/p_0$.  Red values are positive, blue values are negative, and black is zero.  The corresponding animation is included in the supplementary data.  \label{fig:gaussian potential}}
\centering
\includegraphics[width=\wigwid\textwidth]{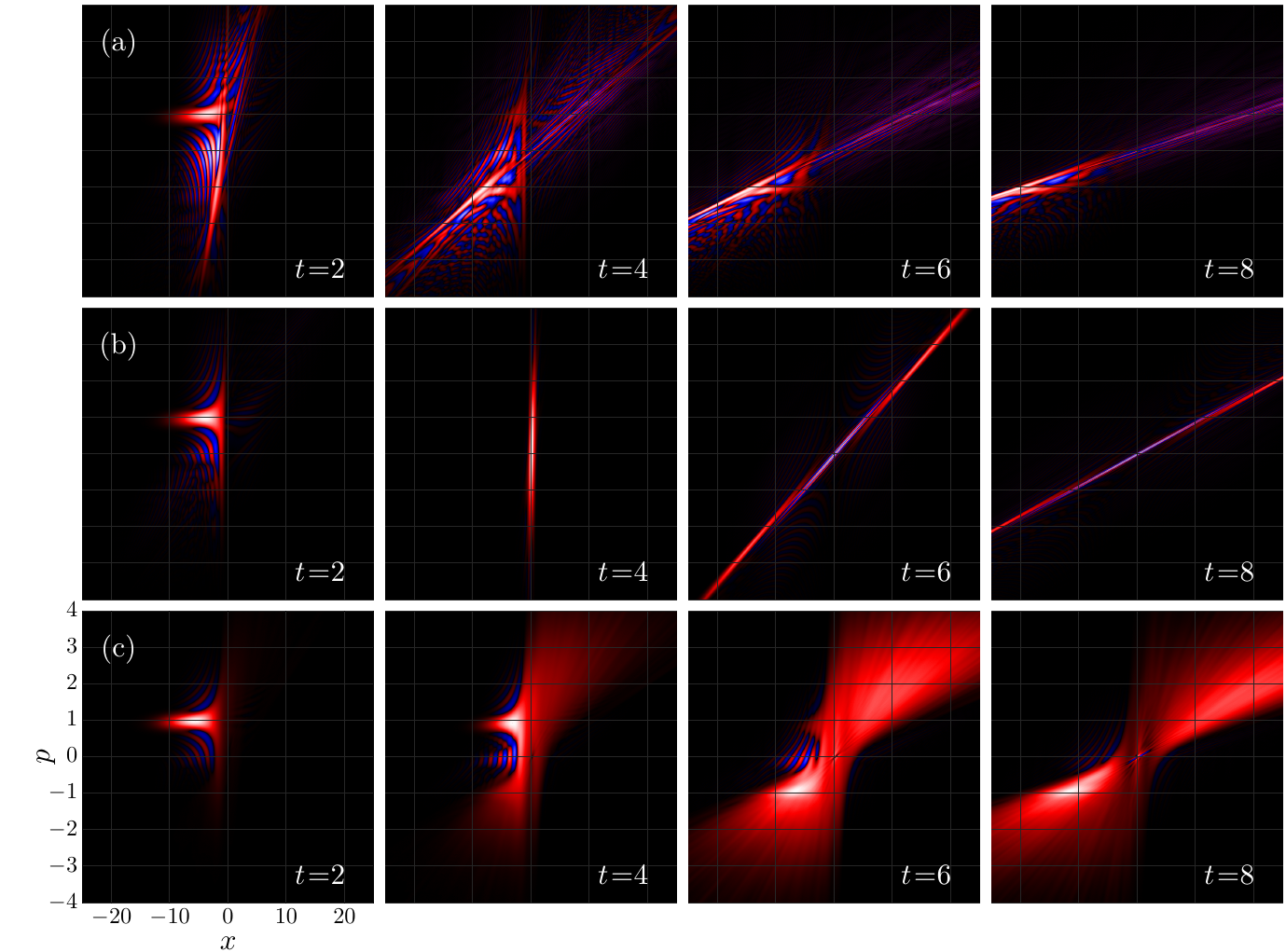}
\caption{Wigner functions for particle incident on a Gaussian measurement with measurement width $\sigma_\mu=1$ and strength  $\kappa=5$; the profile is the same as the potential in Fig.~\ref{fig:gaussian potential}. The initial conditions are the same as in Fig.~\ref{fig:gaussian potential}.  The trajectories are generated with a white-noise-measurement unraveling, as given by the Eq.~(\ref{eq:SME_measurement}).
(a) Single trajectory with reflection from measurement.  
(b) Single trajectory producing a spatial superposition state.   
(c) Ensemble average of measurement.   Red values are positive, blue values are negative, and black is zero.   
The corresponding animations are included in the supplementary data.
\label{fig:gaussian measurement}}
\end{figure*}

Fig.~\ref{fig:gaussian potential} shows a wave packet incident on a Gaussian potential barrier without any measurement.  The wave packet has enough potential energy to cross the barrier (without tunneling).  As with the step potential, the wave packet coherently splits into reflected and transmitted components.  Note that in this case, both parts of the wave packet have the same asymptotic momentum after scattering from the potential barrier.

Fig.~\ref{fig:gaussian measurement}(a) shows a single trajectory where the majority of the wave packet reflects from the Gaussian measurement, without an explicit potential.  Since the Gaussian is a smooth function, the wave packet readily enters regions where the measurement strength is small.   The Gaussian function lacks a sharp edge, and it is possible to get a series of results that weakly localize the atom in the soft edges of the function.  A small part transmits through on this particular trajectory, as indicated by the striations showing the coherence between the reflected and transmitted parts of the wave packet.  

Fig.~\ref{fig:gaussian measurement}(b) shows an alternative case where the wave packet is localized by the Gaussian measurement.  The sides of the Gaussian act as a linear measurement for wave packets confined to one side.  Each side can distinguish different positions within that side and act to localize the wave packet inside the measurement region.  The wave packet then diffuses around from the momentum kicks of the measurement.  This has occurred by $t=4$ in Fig.~\ref{fig:gaussian measurement}(b).  If the wave packet passes over the peak of the Gaussian then it will be driven to a superposition state, since the measurement is locally even in $x$ at the maximum.   If the wave packet is moving quickly then the superposition will tend to be biased in the direction of motion.  The creation of a superposition state is probabilistic, and depends both on localizing the wave packet, and the wave packet diffusing over the peak of the measurement function.  After the superposition forms, the two components diffuse away from the peak of $\mu(x)$.  The symmetry of the measurement ensures that the superposition is not disturbed once it is created.   A superposition state has been created by $t=6$ in Fig.~\ref{fig:gaussian measurement}(b).

Fig.~\ref{fig:gaussian measurement}(c) shows the ensemble average over $10^4$ trajectories for a Gaussian measurement.  Once again, the coherent reflection survives the ensemble average.  The interference fringes between the reflected 
and incident wave packets are easier to see, since the incoherent features from weak localization and partial transmission have averaged out.  Both the reflected and transmitted parts of the wave packet have an incoherent, heated part; this comes from realizations where the particle is localized and split into a superposition state.  In general, the phase of each superposition is random and the superpositions are created at random times, so the coherent features of each superposition are washed out in the ensemble average.  The ensemble-averaged results do not condition on any measurement scheme, and correspond to the unconditioned density operator---the solution to the unconditioned master equation (13).  Thus we will see coherent reflection from \textit{any} process that realizes master equation (13) in the ensemble average.

\subsection{Heating}

We can estimate the momentum transferred by the measurement by considering the Fourier transform of the measurement function.  This is related to the transverse momentum distribution of the bath particles that interact with the particle undergoing the position measurement.  In the resonance-fluorescence example, the bath particles are photons that form the mode profile of the intensity distribution which becomes the measurement function $\mu(x)$.  We obtain the rate of momentum disturbance by multiplying the Fourier transform of $\mu(x)$ by the measurement rate $\kappa$, which is the rate at which we gather position information---and hence disturb momentum.  The approximate final momentum distribution will be the convolution of the particle's momentum distribution with the Fourier transform of the measurement function.  

For example, the Fourier transform of the step function $\theta(x)$ is 
\begin{equation}
\int dk \,\theta(x)e^{ikx} = \pi\delta(k)+i\frac{\mathrm{P.V.}}{k},
\end{equation}
where $\mathrm{P.V.}$ denotes the Cauchy principal value.  The sharp edge of the step implies an arbitrarily fine resolution of how close the particle is to the edge, which implies a divergent momentum disturbance.  The convolution with $1/k$ leads to long tails in momentum space and a divergent momentum uncertainty.  

The convolution of momentum distributions as a result of ``which-way'' measurements has been noted before in the context of projective measurements~\cite{Wiseman97} and transverse momentum transfer.  The convolution result is exact for a projective measurement,  but must incorporate dynamical evolution for continuous measurement.  

The Fourier transform of a Gaussian is also Gaussian, so the momentum transfer in this case is finite.  The width of the momentum distribution scales as the inverse of the width of the measurement resolution.  A narrow measurement function constitutes a locally stronger measurement of position and hence a larger disturbance.  

There is an alternative, measurement-theoretic explanation for the heating.  A strong measurement ensures that we detect
if the particle crosses between regions of differing $\mu(x)$.  Consider the time when the particle begins to cross into the next region, as in the $t=2$ and $t=4$ panels of Fig.~\ref{fig:step_measurement}(b), or the $t=2$ panel of Fig.~\ref{fig:gaussian measurement}(b).  If the measurement record does not yet indicate that the particle has crossed through, then the particle can only have penetrated a small distance into the next region.  Then, when the measurement indicates that the particle has passed into the next region, the state collapses down to the small part that penetrated the next
region.  Since this part is tightly confined in position, there now is a large uncertainty in momentum.  The wave function is effectively multiplied by $\mu(x)$, and so the wave function will acquire the same character in momentum space.    This is apparent, for example, in the $t=4$ frame of Fig.~\ref{fig:gaussian measurement}(b).

\section{Quantum Zeno Effect}
 \label{sec:reflection}
\begin{figure}
\centering
\includegraphics[width=3.38in]{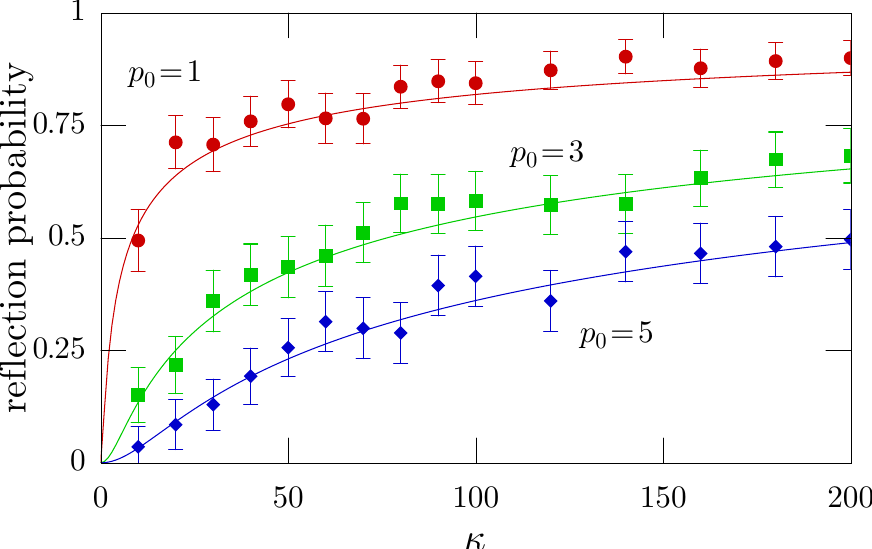}
\caption{Measurement-induced reflection probabilities for a  wave packet with $\sigma_x=10$ incident on $\mu(x)=\theta(x)$.  Numerical simulations (averaged over 256 trajectories) are compared with the analytical result from Eq.~(\ref{eq:Pdet}).}
\label{fig:reflplot}
\end{figure}

Despite the heating, a particle will almost certainly reflect from a strong measurement.  As Fig.~\ref{fig:reflplot} shows, the probability for reflection tends to unity as the measurement strength increases.  This is a manifestation of the quantum Zeno effect~\cite{Allcock69_1, Allcock69_2, Allcock69_3, Misra77,Itano90,Kwiat99,Fischer01}. 

 A simple argument in the case of a step-function measurement $\mu(x) = \theta(x)$, as in Sec.~\ref{section:step-function}, demonstrating this in the limit of a sequence of projective measurements separated by infinitesimal time intervals ($\kappa\longrightarrow \infty$) is as follows~\cite{Echanobe08}.  We assume that at time $t$ the particle is completely confined to the region $x<0$, as enforced by the projective measurement in the immediate past.
The operator for evolution between measurements is
\begin{equation}
U(dt) =\exp\left(-i\frac{H}{\hbar}\,dt\right)= 1-i\frac{H}{\hbar}\,dt.
\end{equation}
The Hamiltonian is responsible for any evolution that could transfer the wave packet into the region $x>0$.  The term proportional to $dt$ is the probability amplitude for the particle to cross into the region $x>0$ during the infinitesimal time step, while the amplitude for staying in the $x<0$ region is unity.  The probability of detecting the particle in the region $x>0$ is thus $O(dt^2)=0$ for the next projective measurement, resetting the particle to the $x<0$ region
and thus confining it there. A similar argument has been made before from the perspective of a ``gap'' in quantum mechanics  ~\cite{Mielnik94}. 

A more quantitative argument can be used in the case of finite measurement strength.  This is related to quantum reflection of atoms dropped at grazing incidence onto ridged surfaces~\cite{Oberst05, Shimizu02}.  This ``quantum reflection'' has been related to the Zeno effect since the ridges act heuristically as absorbers for the atoms, and thus increase the reflection probability from the surface~\cite{Kouznetsov05}.  This treatment misses the measurement back-action and heating that we consider here.  Our calculation here also makes a rigorous connection of this absorber model to continuous-measurement theory.  

Here we will model the evolution by a jump-process realization for $\mu(x)=\theta(x)$,
\begin{eqnarray}\label{eq:SSE_jump}
d|\psi\rangle =& -i\frac{p^2}{2m\hbar}|\psi\rangle \,dt - \kappa\left(\theta^2(x)-\langle\theta^2(x)\rangle\right)|\psi\rangle \,dt \nonumber\\
 & + \left(\frac{\theta(x)|\psi\rangle}{\langle \psi|\theta(x)|\psi\rangle^{1/2}}-|\psi\rangle\right)\,dN,
\end{eqnarray}
where $dN$ is a Poisson process with mean
\begin{equation}
\left<\!\left<dN\right>\!\right> = 2\kappa\langle \theta^2(x)\rangle \,dt.
\end{equation}
This stochastic Schr\"odinger equation gives the same unconditioned evolution as the white-noise unraveling in Eq.~(\ref{eq:SME_measurement}).  We will model the incoming wave packet as a unit-amplitude plane wave, which gives the correct  reflection coefficient as long as the momentum distribution is narrow compared to the mean incident momentum. 
   
We can use a linear trajectory and solve for the between-jump evolution (the nonlinear term represents the probability of the particle to be detected in the region $x>0$, which is precisely what we are computing here).  The linear, between-jump evolution is governed by
\begin{equation}
i\hbar\partial_t|\psi\rangle = \left[ \frac{p^2}{2m}-i\hbar\kappa\theta(x) \right]|\psi\rangle.
\end{equation}
We assume the particle is initially located on the left with mean energy $p_0^2/2m$, and enforce continuity of the wave to match the incoming wave with a reflected wave of the same energy and a decaying solution for $x>0$, which gives the total probability for the particle to be in the ``transmitted'' region.  This is used along with the mean detection rate to find the total detection probability:
\begin{eqnarray}
P_{\mathrm{det}} &= 2\kappa\int\limits_{-\infty}^{\infty} dx\, \theta(x)|\psi(x)|^2 \\
& = \frac{2\sqrt{2}\xi/\chi}{\left(\sqrt{2}+\xi^2/\chi^2\right)^2+\chi^2}\label{eq:Pdet},
\end{eqnarray}
where $\xi := 2m\hbar\kappa/p_0^2$ is the measurement strength scaled in units of the incident kinetic energy, and $\chi := \sqrt{\sqrt{1+\xi^2}-1}$.  For large $\kappa$ the detection probability, and thus the transmission probability, vanishes as $\kappa^{-3/2}$.  

In this linear-trajectory picture, the measurement can be viewed as an imaginary potential that absorbs particles.  The absorbed particles are detected inside the $x>0$ region, and their wave functions are modified according to
\begin{equation}
|\psi\rangle \rightarrow \frac{\theta(x)|\psi\rangle}{\langle \psi|\theta(x)|\psi\rangle^{1/2}}.
\end{equation}
The remainder of the particles coherently reflect from the measurement with a probability given by the norm of the state.  

In Fig.~\ref{fig:reflplot} we have counted only the coherent part of the reflection by computing $\int\limits_{-3\sigma_p}^{+3\sigma_p}dp\, |\phi(-p_0+p)|^2$, where $\phi(p)$ is the momentum-space wave function.  This omits the contribution from the diffusely reflected parts.  As the initial momentum increases,  the particle penetrates farther into the $x>0$ region, and hence the transmission probability increases.   

This calculation works because the step measurement is a two-outcome measurement---as soon as the particle is detected it collapses into the measurement region.   This is analogous to the calculations that can be made for the probability of spontaneous emission from a two-level atom~\cite{Carmichael93}.  This approach can in principle be used to calculate the coherent reflection probabilities given that the particle remains outside the measurement region for a general measurement function.  

Although the reflection probability can approach unity, the ensemble-averaged momentum is conserved, as shown in Eq.~(\ref{eq:mean p}).  The conservation of mean momentum ties together the observation of increasing reflection and the heating of the transmitted wave as $\kappa$ increases.  As $\kappa$ increases the probability for reflection increases as the wave packet is excluded from the measurement region.  Concomitantly, any increase in $\kappa$ increases the heating from the measurement since the evanescent tail in the absorbing region is more tightly confined.  This offsets the decrease in mean momentum from the reflection.  

The Zeno effect has also been noted when combining a continuous position measurement of $\mu(x)=x$ with a  double-well potential~\cite{Gagen93}.  For moderate measurement strengths, the position measurement localizes the particle in one well and inhibits the tunneling between the two wells.  This breaks down for large measurement strengths since the heating from the measurement boosts the energy of particle above the barrier in the middle.  In our case the Zeno effect arises because of the inhomogeneity of the measurement function, rather than the inhomogeneity of the potential.  

\section{Physical  Realizations}

\subsection{Stochastic Dipole Force}
\label{sec:physical realization}
The above effects can be realized by coupling a two-level atom to a high-intensity light field with position-dependent Rabi frequency.  This can be viewed in terms of the stochastic dipole force~\cite{Dalibard85}.   We write the position-dependent Rabi frequency as $\Omega(x)=|\Omega| g(x)$, where $|\Omega| = |\mathbf{d}_{eg}\cdot \mathbf{E}|/\hbar$,  $|\mathbf{E}|$ is the maximum electric field amplitude, and $g(x)$ is a complex, dimensionless mode profile with $|g(x)|\le1$.  The atom obeys
\begin{equation}
\partial_t\rho = -i\left[ \frac{p^2}{2m\hbar}+\Delta\sigma^\dag\sigma+\frac{1}{2}\left[\Omega^*(x)\sigma+\Omega(x)\sigma^\dag\right],\rho\right] + \Gamma\mathcal{D}[\sigma]\rho,
\end{equation}
where $\Gamma$ is the excited state decay rate.  In the high intensity limit, the dressed states $\ket{\pm}$ are approximately equal superpositions of the ground state $\gket$ and the excited state $\eket$, and vice versa.  Each spontaneous-emission event projects the atom into the ground state, and thus into an equal superposition of the dressed states.  The atom therefore sees both shifts $\pm\hbar|\Omega(x)|/2$ of the respective dressed states $\ket{\pm}$.  The interpretation is as follows: after each spontaneous emission event, the atom sees a force
\begin{equation}
 \mathbf{F}=\mp\frac{\hbar\nabla|\Omega|}{2}   ,
\end{equation}
where the sign is chosen randomly, but with equal probability for the two possibilities. Assuming the atom is moving slowly, the momentum change associated with a single spontaneous-emission event is
\begin{equation}
 \Delta p = \mp\frac{\hbar\nabla|\Omega|}{2}   \xi
\end{equation}
where $\xi$ is the time until the next spontaneous emission event, which is a random variable ($\xi>0$) of mean $2/\Gamma$ and exponential probability density
\begin{equation}
 f(\xi) = \frac{\Gamma}{2}\exp\left(-\frac{\Gamma}{2}\xi\right).
\end{equation}
To take into account the randomness of the sign, we can write
\begin{equation}
 \Delta p =  \frac{\hbar\nabla|\Omega|}{2}   \xi'
\end{equation}
where $\xi'\in\mathbb{R}$ has a two-sided exponential probability density
\begin{equation}
 f_\pm(\xi') = \frac{\Gamma}{4}\exp\left(-\frac{\Gamma}{2}|\xi'|\right).
\end{equation}
Then the mean-square kick is
\begin{equation}
 \expct{(\Delta p)^2} =\frac{\hbar^2(\nabla|\Omega|)^2}{4} \,\expct{\xi'^2}
   = \frac{2\hbar^2(\nabla|\Omega|)^2}{\Gamma^2},
\end{equation}
where
\begin{equation}
 \expct{\xi'^2} = \frac{\Gamma}{4}\int_{-\infty}^\infty d\xi'\, \exp\left(-\frac{\Gamma}{2}|\xi'|\right) \xi'^2
 = \frac{8}{\Gamma^2}.
\end{equation}
The diffusion rate is the mean-square step divided by the average step time $\Delta t=2/\Gamma$, so
\begin{equation}
 D_\mathrm{p} = \frac{\expct{(\Delta p)^2}}{\Delta t}
   = \frac{\hbar^2(\nabla|\Omega|)^2}{\Gamma}
   = \frac{\hbar^2|\Omega|^2}{\Gamma} |g'(x)|^2.
   \label{SDFdiffusion}
\end{equation}
This heuristic result agrees with the high-intensity limit of more rigorous but cumbersome calculations ~\cite{Gordon80, Dalibard85, Steck06notes}.

Since the mean time $\Gamma/2$ between changes in the force is much smaller than typical time scales for the motion of cold atoms, we can think of the atomic motion in a temporally coarse-grained sense. Then the stochastic evolution of the atomic motion over a time interval of interest involves many transitions of the potential. The momentum changes have finite variance, so we can invoke the central limit theorem, and treat the force fluctuations as (Gaussian) white noise. An equivalent, Stratonovich stochastic potential that gives rise to the same diffusion rate is thus
\begin{equation}
  V(x,t) = \frac{|\Omega|}{\sqrt{\Gamma}}\sqrt{|g(x)|^2}\circ \frac{dW}{dt},
\end{equation}
where we have dropped the mean dipole potential, which vanishes on resonance $\Delta=0$. We can see that this form of the potential is correct by comparing Eqs.~(\ref{eq:SME_potential}) and (\ref{eq:potential_strat}), so that the Ito form of the SME has the same form as Eq.~(\ref{eq:SME_potential}), with  $\kappa = |\Omega|^2/2\Gamma$ and $\mu(x) = \sqrt{g^\dag(x)g(x)}$, where the sign of $\mu(x)$ is defined such that $|\mu'(x)|^2=|g'(x)|^2$.  We can then read off the momentum-diffusion coefficient using  Eq.~(\ref{kappadiffusioncoeff}) to see that it agrees exactly with Eq.~(\ref{SDFdiffusion}).

It is interesting to note that in this high-intensity regime, the stochastic potential represents an interplay between the  fluctuations in the atomic dipole and the field intensity.  That is, this effect is due to fluctuations in the atom--field interaction Hamiltonian itself.  As such, it does \textit{not} saturate for high intensities, as does spontaneous emission.  

A measurement interpretation of the stochastic dipole force is as follows.  The dipole force for an atom in a particular dressed state arises due to coherent scattering of photons between different wave vectors of the driving field.  In principle, this coherent redistribution of light can be measured to obtain information about the position of the atom in the field, as for example is the case in recoil-induced resonances~\cite{Guo92}.  However, on resonance, the atom flips randomly between the dressed states, so that the \textit{mean} redistribution of photons (and hence the mean dipole force) vanishes.  The measurement information is instead encoded in the \textit{variance} of the redistributed light.

Note that a low-intensity jump decomposition of resonance fluorescence has been considered in the context of realizing imaginary potentials~\cite{Ruschhaupt04}, though without exploring the consequences for the atomic dynamics.  Unfortunately, in the low-intensity limit the spontaneous emission recoil obscures the measurement effects, and this model can break down when pushed to the high intensities required for a large measurement strength.

\subsection{Cavity Quantum Electrodynamics}

A spatially varying position measurement can also be realized in the interaction of a two-level atom with the field of an optical cavity, which is in turn driven by a classical field.  Due to a large detuning, and a large cavity decay rate we can eliminate the atom's internal dynamics and the cavity dynamics, leaving the only the center-of-mass motion of the atom.  In this case the cavity mode function becomes the measurement function~\cite{Doherty98,Steck04,Steck06}.  The simulations from Sec.~\ref{sec:Gaussian} correspond to Gaussian mode of a ring cavity.  

We assume that the cavity is being monitored by homodyne detection, so the system obeys the following master equation \cite{Steck06}
\begin{equation}
d\rho = -\frac{i}{\hbar}[H,\rho]\, dt  +\Gamma\mathcal{D}[\sigma]\rho\, dt + \gamma\mathcal{D}[a]\rho\, dt + \sqrt{\gamma}\mathcal{H}[a]\rho\, dW,
\end{equation}
where $\Gamma$ is the free-space spontaneous emission rate, $\gamma$ is the cavity decay rate, and $a$ is the cavity annihilation operator. The total Hamiltonian is given by
\begin{equation}
H = \frac{\mathbf{p}^2}{2m}+\hbar\omegac a^\dag a+\hbar\omegaa \sigma^\dag\sigma+\hbar g(\mathbf{x})(a^\dag\sigma+a\sigma^\dag)+ \hbar E(a e^{i\omegac t}+a^\dag e^{-i\omegac t}),
\end{equation}
where $\omegac$ is the cavity mode resonance, $\omegaa$ is the atomic transition frequency, $g(\mathbf{x})$ is the cavity mode profile, and $E = \sqrt{\gamma P/\hbar\omegac}$ is the amplitude for the classical driving field, where $P$ is the power in the driving field.  If we transform to a rotating frame, we can remove the free evolution due to the cavity.  Since the classical driving field is resonant with the cavity, we can then write 
\begin{equation}
H = \frac{\mathbf{p}^2}{2m}+\hbar\Delta\sigma^\dag\sigma+\hbar g(\mathbf{x})(a^\dag\sigma+a\sigma^\dag)+ \hbar E(a+a^\dag),
\end{equation}
where $\Delta = \omegaa-\omegac$.   In the limit $\Delta \gg g(\mathbf{x}),\gamma,\Gamma,\mathbf{p}^2/(2m\hbar)$, we can adiabatically eliminate the excited state.  This amounts to replacing $\sigma$ and $\sigma^\dag $ with $\langle \sigma\rangle$ and $\langle \sigma^\dag \rangle$. (For a more rigorous approach see Ref.~\cite{Doherty98}.)  In this limit the effective Hamiltonian is given by 
\begin{equation}
H_{\mathrm{eff}} = \frac{p^2}{2m} + \hbar \frac{g^2(\mathbf{x})}{\Delta} a^\dag a + \hbar E(a + a^\dag),
\end{equation}
and the spontaneous emission terms can be dropped since the effective decay rate is $\mathcal{O}[\Delta^{-2}]$.   If we further assume that the cavity is strongly damped such that $\gamma \gg g^2/\Delta, E, \mathbf{p}^2/(2m\hbar)$, we can also eliminate the cavity.  The final result is the following master equation for the center-of-mass dynamics,
\begin{equation}
d\rho = -\frac{i}{\hbar}[H_{\mathrm{eff}},\rho]\, dt + 2\kappa\mathcal{D}[\mu(\mathbf{x})]\rho\, dt + \sqrt{2\kappa}\mathcal{H}[\mu(\mathbf{x})]\, dW,
\end{equation}
where 
\begin{equation}
\setlength{\arraycolsep}{0ex}
\renewcommand{\arraystretch}{1.9}
\begin{array}{rcl}
H_{\mathrm{eff}} &{}={}&\displaystyle \frac{p^2}{2m} + \hbar \frac{\alpha^2g_0^2}{\Delta}\mu(\mathbf{x})\\
\kappa &{}={}&\displaystyle \frac{\alpha^2g_0^4}{\Delta^2\gamma}\\
\mu(x) &{}={}&\displaystyle \frac{g^2(x)}{g_0^2},
\end{array}
\end{equation}
where $\alpha = 2 E/\gamma$, and $g_0 = \mathrm{max}[g(\mathbf{x})]$.  The mean potential in the effective Hamiltonian can be cancelled by the Stark shift from an off-resonant classical field that does not resonate with a cavity.  This interaction has the form
\begin{equation}
H_{\mathrm{Stark}} = -\hbar\frac{|\Omega(\mathbf{x})|^2}{4\delta}.
\end{equation}
Cancelling the mean potential allows the measurement effects we have discussed to come to the fore.  The master equation then has the same form as Eq.~(\ref{eq:SME_measurement}).  

\subsection{Imaginary Potentials}

Measurements can also be viewed as imaginary (absorbing) potentials, as seen in Sec.~\ref{sec:reflection}.  Imaginary potentials have been used to coherently diffract atomic beams~\cite{Oberthaler96,Oberthaler99} and reshape wave functions to counteract the expansion of a free particle~\cite{Stutzle05}.   A resonant standing wave of light acts like an imaginary potential for two-level atoms.  When the atom absorbs a photon from the resonant standing wave, it will decay to another state, thus being effectively absorbed from the initial beam.  The diffraction and reshaping both rely on post-selecting atoms that have not undergone spontaneous emission.  The surviving atoms are thus most likely to be near the nodes of the standing wave.  For an atomic beam, the standing wave acts as a diffraction grating and enables Bragg diffraction for atomic beams.  This so-called ``anomalous diffraction'' has been further explained in terms of degeneracies of the imaginary Hamiltonian~\cite{BerryOdell98,Berry98}.  The shaping can be viewed as conditioning on null detection of a $\sin^2(k x)$ measurement of position, while the diffraction is a consequence of scattering off the periodic array of nodes.  

\section{Analogies}
\label{sec:analogies}

The coherent reflection from a measurement is analogous to the coherent back-scatter of light from a disordered atomic medium~\cite{Labeyrie99}.  The analogy is clearer from the fluctuating-potential picture, since the paraxial wave equation is the same as the Schr\"odinger equation with time replaced by the propagation direction.  Under the same correspondence, the fluctuating potential in time becomes a disordered potential in space.  In the ensemble average the fluctuating potential also yields a coherent reflection, since it obeys the same unconditioned equations as the measurement.  

An alternative analogy is the reflection of a photon from a conducting surface.  The large complex permittivity---whose imaginary part represents absorption---leads to a rapid extinction of the wave inside the conductor if the conductivity is large.  This is analogous to a large measurement strength for detecting the electromagnetic wave inside the region.   This leads to the large reflection probability for light from the surface of a good conductor, even though it is a good absorber for waves \textit{inside} the medium.

The above examples capture the coherence of the reflection, but miss the transmission aspects of the measurement since the inbound particle is absorbed by the interaction.  Unlike photons, atoms are not destroyed upon detection, so the form of the transmitted wave packet is necessary for a complete picture of measurement-induced reflection.

\section{Outlook}
\label{sec:outlook}
We have outlined a theory of continuous position measurements that describes a spatially varying measurement strength.  The effects shown here require the measurement gradient to dominate over all other dynamics.  It is only in the limit of a large measurement gradient on the scale of the atom, large intensities and small velocities that coherent reflection can take place.   This work can be extended to include other couplings and position measurements such as atoms coupled to micro-toroidal resonators~\cite{Spillane05}, and the effect of measurements when the particle is confined to a potential.  The case of both an inhomogeneous measurement and potential is of interest, for example, for studying the quantum--classical transition for classically chaotic systems.  Future work will extend the idealized theory presented here to account for the imaging setup, diffraction effects and realistic photo-detection.  

\section{Acknowledgments}

The authors wish to acknowledge discussions with Tanmoy Bhattacharya, Robin Blume-Kohout, Steven van Enk, and Michael Raymer, and critical readings of the manuscript by Eryn Cook, Paul Martin, Elizabeth Schoene, and Jeremy Thorn.  JBM and DAS are  supported by the National Science Foundation, under Project No. PHY-0547926.  KJ is supported by the National Science Foundation under Project No. PHY-0902906.


\begin{thebibliography}{10}

\bibitem{Rocheleau10}
T.~Rocheleau, T.~Ndukum, C.~Macklin, J.~B. Hertzberg, A.~A. Clerk, and K.~C.
  Schwab.
\newblock Preparation and detection of a mechanical resonator near the ground
  state of motion.
\newblock {\em Nature}, 463:72, 2010.

\bibitem{Murch08}
Kater~W. Murch, Kevin~L. Moore, Subhadeep Gupta, and Dan~M. Stamper-Kurn.
\newblock Observation of quantum-measurement backaction with an ultracold
  atomic gas.
\newblock {\em Nature Physics}, 4:561, 2008.

\bibitem{Belavkin87}
V.~P. Belavkin.
\newblock Non-demolition measurement and control in quantum dynamical systems.
\newblock In A.~Blaquiere, S.~Diner, and G.~Lochak, editors, {\em Information,
  Complexity and Control in Quantum Physics}, page 331. Proceedings of the 4th
  International Seminar on Mathematical Theory of Dynamical Systems and
  Microphysics, Sept.\ 1985, Springer-Verlag, 1987.

\bibitem{Wiseman93}
H.~M. Wiseman and G.~J. Milburn.
\newblock Quantum theory of field-quadrature measurements.
\newblock {\em {Phys.\ Rev.\ A}}, 47:642, 1993.

\bibitem{Caves87}
Carlton~M. Caves and G.~J. Milburn.
\newblock Quantum-mechanical model for continuous position measurements.
\newblock {\em {Phys.\ Rev.\ A}}, 36:5543, 1987.

\bibitem{Gagen93}
M.~J. Gagen, H.~M. Wiseman, and G.~J. Milburn.
\newblock Continuous position measurements and the quantum {Z}eno effect.
\newblock {\em {Phys.\ Rev.\ A}}, 48:132, 1993.

\bibitem{Holland96}
M.~Holland, S.~Marksteiner, P.~Marte, and P.~Zoller.
\newblock Measurement induced localization from spontaneous decay.
\newblock {\em {Phys.\ Rev.\ Lett.}}, 76(20):3683, 1996.

\bibitem{Jacobs06}
Kurt Jacobs and Daniel.~A. Steck.
\newblock A straightforward introduction to continuous quantum measurement.
\newblock {\em {Cont.\ Phys.}}, 47:279, 2006.

\bibitem{Doherty98}
A.~C. Doherty, A.~S. Parkins, S.~M. Tan, and D.~F. Walls.
\newblock Motional states of atoms in cavity {QED}.
\newblock {\em {Phys.\ Rev.\ A}}, 57:4804, 1998.

\bibitem{Hood00}
C.~J. Hood, T.~W. Lynn, A.~C. Doherty, A.~S. Parkins, and H.~J. Kimble.
\newblock The atom-cavity microscope: Single atoms bound in orbit by single
  photons.
\newblock {\em Science}, 287(5457):1447, 2000.

\bibitem{Doherty99}
A.~C. Doherty and K.~Jacobs.
\newblock Feedback control of quantum systems using continuous state
  estimation.
\newblock {\em {Phys.\ Rev.\ A}}, 60(4):2700, 1999.

\bibitem{Fischer02}
T.~Fischer, P.~Maunz, P.~W.~H. Pinkse, T.~Puppe, and G.~Rempe.
\newblock Feedback on the motion of a single atom in an optical cavity.
\newblock {\em {Phys.\ Rev.\ Lett.}}, 88(16):163002, 2002.

\bibitem{Steck04}
Daniel~A. Steck, Kurt Jacobs, Hideo Mabuchi, Tanmoy Bhattacharya, and Salman
  Habib.
\newblock Quantum feedback control of atomic motion in an optical cavity.
\newblock {\em {Phys.\ Rev.\ Lett.}}, 92(22):223004, 2004.

\bibitem{Steck06}
Daniel~A. Steck, Kurt Jacobs, Hideo Mabuchi, Salman Habib, and Tanmoy
  Bhattacharya.
\newblock Feedback cooling of atomic motion in cavity {QED}.
\newblock {\em {Phys.\ Rev.\ A}}, 74(1):012322, 2006.

\bibitem{Kubanek09}
A.~Kubanek, M.~Koch, C.~Sames, A.~Ourjoumtsev, P.~W.~H. Pinkse, K.~Murr, and
  G.~Rempe.
\newblock Photon-by-photon feedback control of a single-atom trajectory.
\newblock {\em Nature}, 462:898, 2009.

\bibitem{Steixner05}
V.~Steixner, P.~Rabl, and P.~Zoller.
\newblock Quantum feedback cooling of a single trapped ion in front of a
  mirror.
\newblock {\em {Phys.\ Rev.\ A}}, 72(4):043826, 2005.

\bibitem{Bushev06}
Pavel Bushev, Daniel Rotter, Alex Wilson, Fran{\c{c}}ois Dubin, Christoph
  Becher, J\"urgen Eschner, Rainer Blatt, Viktor Steixner, Peter Rabl, and
  Peter Zoller.
\newblock Feedback cooling of a single trapped ion.
\newblock {\em {Phys.\ Rev.\ Lett.}}, 96(4):043003, 2006.

\bibitem{Hopkins03}
Asa Hopkins, Kurt Jacobs, Salman Habib, and Keith Schwab.
\newblock Feedback cooling of a nanomechanical resonator.
\newblock {\em {Phys.\ Rev.\ B}}, 68(23):235328, 2003.

\bibitem{Bhattacharya05}
Tanmoy Bhattacharya, Salman Habib, and Kurt Jacobs.
\newblock Continuous quantum measurement and the emergence of classical chaos.
\newblock {\em {Phys.\ Rev.\ Lett.}}, 85:4852, 2000.

\bibitem{Habib06}
Salman Habib, Kurt Jacobs, and Kosuke Shizume.
\newblock Emergence of chaos in quantum systems far from the classical limit.
\newblock {\em {Phys.\ Rev.\ Lett.}}, 96:010403, 2006.

\bibitem{Allcock69_1}
G.~R Allcock.
\newblock The time of arrival in quantum mechanics {I}. formal considerations.
\newblock {\em {Ann.\ Phys. (N.Y.)}}, 53:253, 1969.

\bibitem{Allcock69_2}
G.~R Allcock.
\newblock The time of arrival in quantum mechanics {II}. the individual
  measurement.
\newblock {\em {Ann.\ Phys.\ (N.Y.)}}, 53:286, 1969.

\bibitem{Allcock69_3}
G.~R Allcock.
\newblock The time of arrival in quantum mechanics {III}. the measurement
  ensemble.
\newblock {\em {Ann.\ Phys.\ (N.Y.)}}, 53:311, 1969.

\bibitem{Misra77}
B.~Misra and E.~C.~G. Sudarshan.
\newblock The {Z}eno's paradox in quantum theory.
\newblock {\em {J.\ Math.\ Phys.}}, 18:756, 1977.

\bibitem{Facchi01}
P.~Facchi, S.~Pascazio, A.~Scardicchio, and L.~S. Schulman.
\newblock {Z}eno dynamics yields ordinary constraints.
\newblock {\em {Phys.\ Rev.\ A}}, 65:012108, 2001.

\bibitem{Facchi04}
P.~Facchi, G.~Marmo, S.~Pascazio, A.~Scardicchio, and E.~C.~G. Sudarshan.
\newblock Quantum {Z}eno dynamics.
\newblock {\em {Phys.\ Lett.\ A}}, 6:S492, 2004.

\bibitem{Exner05}
P.~Exner and T.~Ichinose.
\newblock A product formula related to quantum {Z}eno dynamics.
\newblock {\em Ann.\ Henri Poincare}, 6:195, 2005.

\bibitem{Koshino05}
K.~Koshino and A.~Shimizu.
\newblock Quantum {Z}eno effect by general measurements.
\newblock {\em {Phys.\ Rep.}}, 412:191, 2005.

\bibitem{Echanobe08}
J.~Echanobe, A.~del Campo, and J.~G. Muga.
\newblock Disclosing hidden information in the quantum {Z}eno effect: Pulsed
  measurement of the quantum time of arrival.
\newblock {\em {Phys.\ Rev.\ A}}, 77(3):032112, 2008.

\bibitem{raizen2005}
M.~G. Raizen, A.~M. Dudarev, Qian Niu, and N.~J. Fisch.
\newblock Compression of atomic phase space using an asymmetric one-way
  barrier.
\newblock {\em {Phys.\ Rev.\ Lett.}}, 94(5):053003, 2005.

\bibitem{ruschhaupt2004}
A.~Ruschhaupt and J.~G. Muga.
\newblock Atom diode: A laser device for a unidirectional transmission of
  ground-state atoms.
\newblock {\em {Phys.\ Rev.\ A}}, 70(6):061604(R), 2004.

\bibitem{price2008}
G.~N. Price, S.~T. Bannerman, K.~Viering, E.~Narevicius, and M.~G. Raizen.
\newblock Single-photon atomic cooling.
\newblock {\em {Phys.\ Rev.\ Lett.}}, 100(9):093004, 2008.

\bibitem{thorn2008}
J.~J. Thorn, E.~A. Schoene, T.~Li, and D.~A. Steck.
\newblock Experimental realization of an optical one-way barrier for neutral
  atoms.
\newblock {\em {Phys.\ Rev.\ Lett.}}, 100(24):240407, 2008.

\bibitem{Wiseman10QMAC}
H.~M. Wiseman and G.~J. Milburn.
\newblock {\em Quantum Measurement and Control}.
\newblock Cambridge University Press, 2010.

\bibitem{JacobsSP}
K.~Jacobs.
\newblock {\em Stochastic Processes for Physicists: Understanding Noisy
  Systems}.
\newblock CUP, Cambridge, 2010.

\bibitem{GardinerSM}
C.~W. Gardiner.
\newblock {\em Stochastic Methods}.
\newblock Springer, fourth edition, 2009.

\bibitem{Milburn96}
G.~J. Milburn.
\newblock Classical and quantum conditional statistical dynamics.
\newblock {\em {Quant.\ Semiclass.\ Opt.}}, 8, 1996.

\bibitem{Goetsch94}
P.~Goetsch and R.~Graham.
\newblock Linear stochastic wave equations for continuously measured quantum
  systems.
\newblock {\em {Phys.\ Rev.\ A}}, 50:5242, 1994.

\bibitem{Gardiner85}
C.~W. Gardiner and M.~J. Collett.
\newblock Input and output in damped quantum systems: Quantum stochastic
  differential equations and the master equation.
\newblock {\em {Phys.\ Rev.\ A}}, 31(6):3761, 1985.

\bibitem{stochastic_footnote}
Note that a more general system-bath interaction has the form $H_\mathrm{SB} =
  a^\dagger dB_\mathrm{in} + a\, dB_\mathrm{in}^\dagger$, rather than
  $H_\mathrm{SB} = \mu(x) (dB_\mathrm{in} + dB_\mathrm{in}^\dagger)$. However,
  a more general interaction necessarily includes dissipative coupling (i.e.,
  coupling to momentum). The interaction we consider is general for those
  corresponding strictly to \textit{position} measurements.

\bibitem{Dalibard85}
J~Dalibard and C~Cohen-Tannoudji.
\newblock Dressed-atom approach to atomic motion in laser light: the dipole
  force revisited.
\newblock {\em {J.\ Opt.\ Soc.\ Am.\ B}}, page 1707, 1985.

\bibitem{Jacobs09}
Kurt Jacobs, Lin Tian, and Justin Finn.
\newblock Engineering superposition states and tailored probes for
  nanoresonators via open-loop control.
\newblock {\em {Phys.\ Rev.\ Lett.}}, 102:057208, 2009.

\bibitem{Schleich_phase_space}
W.~P Schleich.
\newblock {\em Quantum Optics in Phase Space}.
\newblock Wiley, 2001.

\bibitem{Kwiat99}
P.~G. Kwiat, A.~G. White, J.~R. Mitchell, O.~Nairz, G.~Weihs, H.~Weinfurter,
  and A.~Zeilinger.
\newblock High-efficiency quantum interrogation measurements via the quantum
  {Z}eno effect.
\newblock {\em {Phys.\ Rev.\ Lett.}}, 83(23):4725, 1999.

\bibitem{Paul96}
H.~Paul and Mladen Pavi\v{c}i\'c.
\newblock Resonance interaction-free measurement.
\newblock {\em {Int.\ J.\ Theor.\ Phys.}}, 35:2085, 1996.

\bibitem{Tsegaye98}
T.~Tsegaye, E.~Goobar, A.~Karlsson, G.~Bj\"ork, M.~Y. Loh, and K.~H. Lim.
\newblock Efficient interaction-free measurements in a high-finesse
  interferometer.
\newblock {\em {Phys.\ Rev.\ A}}, 57(5):3987, 1998.

\bibitem{Wiseman97}
H.~M. Wiseman, F.~E. Harrison, M.~J. Collett, S.~M. Tan, D.~F. Walls, and R.~B.
  Killip.
\newblock Nonlocal momentum transfer in \textit{welcher Weg} measurements.
\newblock {\em {Phys.\ Rev.\ A}}, 56:55, 1997.

\bibitem{Itano90}
W.~M. Itano, D.~J. Heinzen, J.~J. Bollinger, and D.~J. Wineland.
\newblock Quantum {Z}eno effect.
\newblock {\em {Phys.\ Rev.\ A}}, 41:2295, 1990.

\bibitem{Fischer01}
M.~C. Fischer, B.~Guti{\'e}rrez-Medina, and M.~G. Raizen.
\newblock Observation of the quantum {Z}eno and anti-{Z}eno effects in an
  unstable system.
\newblock {\em {Phys.\ Rev.\ Lett.}}, 87:040402, 2001.

\bibitem{Mielnik94}
Bogdan Mielnik.
\newblock The screen problem.
\newblock {\em Foundations of Physics}, 24:1113, 1994.

\bibitem{Oberst05}
Hilmar Oberst, Dimitrii Kouznetsov, Kazuko Shimizu, Jun-ichi Fujita, and Fujio
  Shimizu.
\newblock Fresnel diffraction mirror for an atomic wave.
\newblock {\em {Phys.\ Rev.\ Lett.}}, 94(1):013203, 2005.

\bibitem{Shimizu02}
Fujio Shimizu and Jun-ichi Fujita.
\newblock Reflection-type hologram for atoms.
\newblock {\em {Phys.\ Rev.\ Lett.}}, 88(12):123201, 2002.

\bibitem{Kouznetsov05}
Dmitrii Kouznetsov and Hilmar Oberst.
\newblock Reflection of waves from a ridged surface and the {Z}eno effect.
\newblock {\em Optical Review}, 12:363, 2005.

\bibitem{Carmichael93}
H.~J. Carmichael.
\newblock {\em An Open Systems Approach to Quantum Optics}.
\newblock Springer-Verlag, 1993.

\bibitem{Gordon80}
J.~P. Gordon and A.~Ashkin.
\newblock Motion of atoms in a radiation trap.
\newblock {\em {Phys.\ Rev.\ A}}, 21(5):1606, 1980.

\bibitem{Steck06notes}
Daniel~A. Steck.
\newblock Quantum and atom optics.
\newblock course notes available online at \texttt{http://steck.us/teaching},
  2006.

\bibitem{Guo92}
J.~Guo, P.~R. Berman, B.~Dubetsky, and G.~Grynberg.
\newblock Recoil-induced resonances in nonlinear spectroscopy.
\newblock {\em {Phys.\ Rev.\ A}}, 46:1426, 1992.

\bibitem{Ruschhaupt04}
A.~Ruschhaupt, J.~A. Damborenea, B.~Navarro, J.~G. Muga, and G.~C. Hegerfeldt.
\newblock Exact and approximate complex potentials for modelling time
  observables.
\newblock {\em {Europhys.\ Lett.}}, 67(1):1, 2004.

\bibitem{Oberthaler96}
Markus~K. Oberthaler, Roland Abfalterer, Stefan Bernet, J\"org Schmiedmayer,
  and Anton Zeilinger.
\newblock Atom waves in crystals of light.
\newblock {\em {Phys.\ Rev.\ Lett.}}, 77:4980, 1996.

\bibitem{Oberthaler99}
M.~K. Oberthaler, R.~Abfalterer, S.~Bernet, C.~Keller, J.~Schmiedmayer, and
  A.~Zeilinger.
\newblock Dynamical diffraction of atomic matter waves by crystals of light.
\newblock {\em {Phys.\ Rev.\ A}}, 60:456, 1999.

\bibitem{Stutzle05}
R.~St\"utzle, M.~C. G\"obel, Th. H\"orner, E.~Kierig, I.~Mourachko, M.~K.
  Oberthaler, M.~A. Efremov, M.~V. Fedorov, V.~P. Yakovlev, K.~A.~H. van
  Leeuwen, and W.~P. Schleich.
\newblock Observation of nonspreading wave packets in an imaginary potential.
\newblock {\em {Phys.\ Rev.\ Lett.}}, 95:110405, 2005.

\bibitem{BerryOdell98}
M~V Berry and D~H~J O'Dell.
\newblock Diffraction by volume gratings with imaginary potentials.
\newblock {\em {J.\ Phys.\ A}}, 31:2093, 1998.

\bibitem{Berry98}
M~V Berry.
\newblock Lop-sided diffraction by absorbing crystals.
\newblock {\em {J.\ Phys.\ A}}, 31:3493, 1998.

\bibitem{Labeyrie99}
G.~Labeyrie, F.~de~Tomasi, J.-C. Bernard, C.~A. M\"uller, C.~Miniatura, and
  R.~Kaiser.
\newblock Coherent backscattering of light by cold atoms.
\newblock {\em {Phys.\ Rev.\ Lett.}}, 83(25):5266, 1999.

\bibitem{Spillane05}
S.~M. Spillane, T.~J. Kippenberg, K.~J. Vahala, K.~W. Goh, E.~Wilcut, and H.~J.
  Kimble.
\newblock Ultrahigh-{Q} toroidal microresonators for cavity quantum
  electrodynamics.
\newblock {\em {Phys.\ Rev.\ A}}, 71(1):013817, 2005.

\end{thebibliography}
\end{document}